\documentclass[twocolumn,aps,floatfix,showpacs,prl]{revtex4-1}
\usepackage{amsmath,amssymb}
\usepackage{graphicx}
\usepackage{psfrag}
\usepackage[T1]{fontenc}
\usepackage{color}

\begin{document}

\title{Dipolar Molecules in Optical Lattices}
\author{Tomasz Sowi\'nski$^{1,2}$, Omjyoti Dutta$^2$, Philipp Hauke$^2$, Luca Tagliacozzo$^2$, Maciej Lewenstein$^{2,3}$}
\affiliation{
\mbox{$^1$Institute of Physics of the Polish Academy of Sciences, Al. Lotnik\'ow 32/46, 02-668 Warsaw, Poland}\\
\mbox{$^2$ ICFO --The Institute of Photonic Sciences, Av. Carl Friedrich Gauss, num. 3, 08860 Castelldefels (Barcelona), Spain  }
\mbox{$^3$ ICREA -- Instituci{\'o} Catalana de Recerca i Estudis Avan\c{c}ats, Lluis Companys 23, E-08010 Barcelona, Spain} }
\date{\today}
\begin{abstract}
We study the extended Bose--Hubbard model describing an ultracold
gas of dipolar molecules in an optical lattice, taking into account
all on-site and nearest-neighbor interactions, including
occupation-dependent tunneling and pair tunneling terms. Using exact
diagonalization and the multiscale entanglement renormalization
ansatz, we show that these terms can destroy insulating
phases and lead to novel quantum phases. These considerable changes
of the phase diagram have to be taken into account in upcoming
experiments with dipolar molecules.
\end{abstract}
\pacs{37.10Jk,67.85.Hj,75.40.Cx}
\maketitle


Trapping and manipulating ultracold gases in optical lattices has
allowed the realization of many-body physics in a controlled
environment. For atoms interacting via contact interaction, a
quantum phase transition from a superfluid (SF) to a Mott insulator (MI)
has been predicted and observed \cite{Bloch}. In the simplest case,
these systems can be theoretically described by the Bose-Hubbard
(BH) model, which has two parameters: a tunneling $J$ and an on-site
interaction $U$ \cite{Fisher, Jaksch}. A natural extension of the
Bose--Hubbard model comes from including long-range interactions
between particles. Experiments on ultracold polar molecules have
renewed interest in extended Bose-Hubbard models which can model
such systems in optical lattices \cite{Jin1, Jin2,Inoye, RbCs}. Because of
the strong electric dipole moment of polar molecules, long-range
interactions play a crucial role in the collective behavior of the
system, leading to the appearance of states with long-range order,
like various structured insulating states, supersolids, Wigner
crystals, pair-supersolids, etc.\ \cite{Lew1, Lew2, Zol1, Zol2, Dem,
Santos, Lew3}.

In this Letter, we study the ground-state of dipolar molecules in a
2D square optical lattice with a harmonic trapping along the
polarization direction of the dipoles. We derive a modified BH model
which includes additional occupation-dependent nearest-neighbor
(NN) hopping processes arising from long-range dipolar interactions
in the lowest Bloch band. Usually, interaction-induced hopping terms
are neglected when discussing dipolar bosonic molecules.
In this Letter, we show that these terms considerably change the
physics of dipolar soft-core bosons. Soft-core bosons in square and
one-dimensional lattices have been discussed in the literature within
the extended Hubbard model, focusing on the presence of stable
supersolidity \cite{Sen, batr}. In the usual case with only NN
interaction, at sufficient dipolar strength, the ground states at
half- and unit-filling are checkerboard (CB) insulating states.
Using exact diagonalization (ED) and multiscale entanglement renormalization
ansatz (MERA), we solve the one-dimensional extended Hubbard model including the
novel occupation-dependent NN hopping processes. We find that with
increasing dipolar interaction, the system enters from the CB phases
to a novel state which has a one-particle superfluid (SF) and
pair-superfluid (PSF) properties. Particularly we find a region where both of them coexists with
the SF order parameter has alternating sign at consecutive sites.

Our system consists of dipolar bosons polarized by an external
electric field along the $z$ direction and confined in a square optical lattice. The
corresponding Hamiltonian reads
$H=\int\!\!\mathrm{d}^3\boldsymbol{r}\,\,
\Psi^\dagger(\boldsymbol{r}) \left [- \frac{\hbar^2}{2m} \nabla^2 +
V_{\rm latt}(\boldsymbol{r}) \right]\Psi(\boldsymbol{r}) +
\frac{1}{2}\int\!\!\int\mathrm{d}^3\boldsymbol{r}\,\mathrm{d}^3\boldsymbol{r}'
\Psi^\dagger(\boldsymbol{r})\Psi^\dagger(\boldsymbol{r}'){\cal
V}(\boldsymbol{r}-\boldsymbol{r}')
\Psi(\boldsymbol{r})\Psi(\boldsymbol{r}')$, where
$\Psi^\dagger(\boldsymbol{r})$ $(\Psi(\boldsymbol{r})$) are the
bosonic creation (annihilation) field operators. $V_{\rm
latt}(\boldsymbol{r})=V_0 \left [ \sin^2 \frac{2\pi}{\lambda}x +
\sin^2 \frac{2\pi}{\lambda}y \right] + m\Omega_z^2 z^2/2$ is an
external lattice potential of lattice depth $V_0$, generated by a
laser field of wave-length $\lambda$, with $\Omega_z$ characterizing
the external harmonic potential in $z$ direction. The dipole--dipole
interaction is denoted by ${\cal V}(\boldsymbol{r})$. By expanding
the field operator $\Psi(\boldsymbol{r})=\sum_i {\cal
W}_i(x,y)\mathrm{e}^{-\kappa z^2/2}\,\,\hat{a}_i$ in lowest
Bloch-band Wannier-functions ${\cal W}_i(x,y)$, and by restricting
ourselves to on-site and NN terms, we arrive at the extended BH model
\begin{align} \label{Hamiltonian}
H &= -J\sum_{\{ij\}}\hat{a}^\dagger_i\hat{a}_j  + \frac{U}{2}\sum_i\hat{n}_i (\hat{n}_i-1) + V\sum_{\{ij\}}\hat{n}_i \hat{n}_j  \nonumber \\
& - T\sum_{\{ij\}}
\hat{a}^\dagger_i\left(\hat{n}_i+\hat{n}_j\right)\hat{a}_j +
\frac{P}{2}\sum_{\{ij\}}
\hat{a}^\dagger_i\hat{a}^\dagger_i\hat{a}_j\hat{a}_j,
\end{align}
where $\hat{a}_i$ ($\hat{a}^\dagger_i$) annihilates (creates) a
particle on lattice site $i$, $\hat{n}_i =
\hat{a}^\dagger_i\hat{a}_i$ is the corresponding density operator,
$J$ the standard tunneling coefficient, $U$ the on-site interaction, and 
$V$ the NN interaction, arising from a truncation of the dipolar interactions to the dominating term.
Dipolar interactions lead
to two novel terms in Eq.~\eqref{Hamiltonian}: The term proportional
to $T$ describes one-particle tunneling to a neighboring site
induced by the occupation of that site, and the term proportional to $P$ is
responsible for NN pair tunneling \cite{Gorshkov, Illuminati, UFisher}.

The matrix elements $U$, $V$, $T$, and $P$ are given by a sum of
dipolar and $\delta$-like contact interactions,
${\cal V}(\boldsymbol{r}-\boldsymbol{r}')= \left[g\,\delta^{(3)}(\boldsymbol{r}-\boldsymbol{r}') + \gamma \left(\frac{1}{|\boldsymbol{r}-\boldsymbol{r}'|^3}-3\frac{(z-z')^2}{|\boldsymbol{r}-\boldsymbol{r}'|^5}\right)\right]$.
We measure all lengths in units of the laser wave length $\lambda$
and all energies in recoil energies $E_R =
2\pi^2\hbar^2/(m\lambda^2)$, where $m$ is the bosonic mass.
Additionally, we define the lattice flattening $\kappa=\hbar\Omega_z/2E_R$
as well as the dimensionless coupling constants describing contact and dipolar
interaction,
$g = 16\pi^2 a_s/\lambda$ and $\gamma = m
d^2 /(\hbar^2 \varepsilon_0\lambda)$ (where $a_s$ is the s-wave
scattering length, $\varepsilon_0$ is the vacuum permittivity, and
$d$ is the electric dipole moment of the bosons).

\begin{figure}
\centering
\includegraphics[scale=0.8]{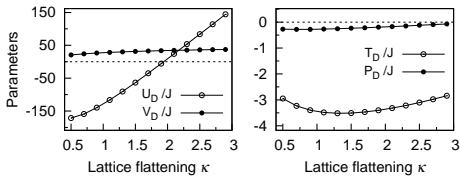}
\caption{ Dependence of the dipolar part (subscript $D$) of $U$,
$V$, $T$, and $P$ on the lattice flattening $\kappa$ for lattice
depth $V_0 = 6E_R$ and $\gamma=52$. }\label{Fig1}
\end{figure}

 For concreteness, we consider an ultracold gas of
dipolar molecules confined in a optical lattice with lattice depth
$V_0 = 6 E_R$, mass $m=220a.m.u$ and $\lambda = 790\,\mathrm{nm}$
\cite{Kotochigova}. We also assume that the s-wave scattering length
of the molecules, $a_s \approx 100a_0$. For these parameters, $g
\approx 1.06$ is approximately constant. We consider dipole moments
$d$ up to $\sim 3\,\mathrm{D}$ ($\gamma$ up to $\sim 470$), which
can be achievable for molecules like bosonic $\mathrm{RbCs},
\mathrm{KLi}$ \cite{KLi} etc.
To illustrate the relative strengths of different parameters, in Fig.\ \ref{Fig1}, we compare for $\gamma = 52$ the tunneling $J$
with the dipolar contribution (subscript $D$) to the parameters $U$,
$V$, $T$ and $P$. For the parameters chosen, $T_{\mathrm{D}}$ and
$P_{\mathrm{D}}$ are 1 orders of magnitude smaller than
$V_{\mathrm{D}}$ where as $U_{\mathrm{D}}/T_{\mathrm{D}}$ can be
tuned by changing $\kappa$. On the other hand, $T_{\mathrm{D}}$ can
dominate over $J$ for large $\gamma$. In addition, $T$ and $J$ can have opposite sign as seen in Fig.\ \ref{Fig1}.  For concreteness, we
choose the lattice parameter $\kappa\approx 1.95$, making
(additionally to $J$) the on-site interaction $U$ almost independent of the
dipole moment ($U_{\mathrm{D}} \approx 0$). In this case, for large
enough $\gamma$, we expect that with increasing $d$ the parameters
$V$, $T$ and $P$ determine the system properties. 
For clarity, we
restrict ourselves to a 1D chain of $N$ lattice sites with periodic
boundary conditions.

\begin{figure}
\centering
\includegraphics[scale=.9]{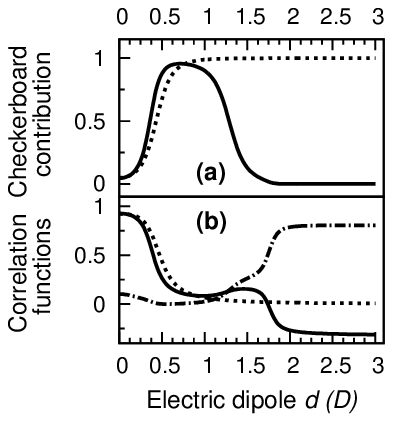} 
\includegraphics[scale=.55]{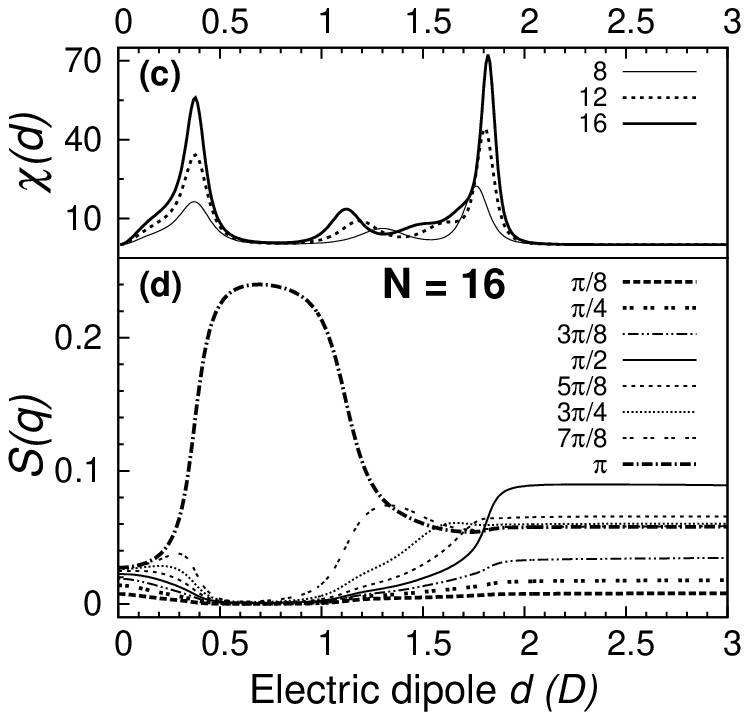}
\caption{We plot various properties of the exact ground state of a
half-filled system as a function of dipole moment $d$. Fig.~(a) shows the
contribution of the CB states to the ground state of the system for $N=8$. 
In Fig.~(b) we plot the one-particle and two-particle correlation functions
$\phi_i$ and $\Phi_i$. The dotted line shows $\phi_i$ when
we neglect the terms $T$ and $P$. When $T,P \neq 0$, the solid line
and dash-dotted line shows $\phi_i$ and $\Phi_i$ respectively as a
function of dipole moment $d$. In Fig.~(c) we
plot the fidelity susceptibility $\chi(d)$ for the half-filled
system for different system sizes. In Fig.~(d) we have shown the
structure-factor $S$ at different ordering wave vectors for the
half-filled system with $16$ sites. }\label{Fig2}
\end{figure}

To get a first understanding of the system, we find the ground state $|\psi_0(d)\rangle$
as a function of $d$ by exact diagonalization (ED) of a half-filled system
with $N=8$ sites. 
We also present results for $N=12$ and $16$ to check for dependence
on system size. Without the occupation-dependent tunneling terms $T$
and $P$, we observe the usual scenario with only two phases, a
single-particle SF and a CB phase. The transition happens at
$d\approx0.4\,\mathrm{D}$. It is marked by an increase of the
contribution of the checkerboard states to the ground state to
almost $100\%$ [inset of Fig.~\ref{Fig2}(a)]. Also, the one-particle
correlation function $\phi_i = \sum_{\{j\}}\langle a^\dagger_j a_i
\rangle$ almost vanishes, indicating the
transition to an insulating state. In the half-filled system, the
transition occurs because for large enough $V$ the particles can
decrease their energy by avoiding every second site. If we neglect
$T$ and $P$, the situation will not change by further increasing $d$
(dotted lines in Fig.\ \ref{Fig2}), since this only increases $V$
even more.
However, the situation changes significantly when we take into
account the density-induced tunneling $T$ and the pair tunneling
$P$. In this case, for $d\approx 1.1\,\mathrm{D}$, a
second phase transition occurs, destroying the CB order [solid lines
in Fig.\ \ref{Fig2}(a)]. Previous studies have completely
neglected such a possible destruction of CB order at large $d$. At
the transition, the contribution of the CB state to the ground state
decreases rapidly, and the one-particle as well as the two-particle
NN correlation function $\Phi_i = \sum_{\{j\}} \langle
a^\dagger_ja^\dagger_j a_i a_i\rangle$ [dashed-dotted line in Fig.\
\ref{Fig2}(b)] attain finite positive values, indicating that the
new phase shows single-particle as well as pair superfluidity. In this region we also find that
the long-ranged correlation function $\langle
a^\dagger_j a_i\rangle$ for $|i-j|\le6$ decays slowly with alternation sign for consecutive sites.
This suggests appearance of antiferromagnetic like order due to the positive hopping $T$ resulting in the condensation of bosons
at the edge of the Brillouin zone. We also looked into the relative effect of $T$ and $P$ on the PSF state. We found that
PSF is generated due to the correlated tunneling term $T$ (in interplay with the nearest-neighbor interaction $V$).  

       For even larger electric moments, a third phase transition happens,
where $\phi_i$ changes sign. Another signature of this
transition is a rapid growth of $\Phi_i$. Since this quantity
measures fluctuations of bosonic pairs, this is a signature of a
novel pair-superfluid (PSF) phase. The appearance of pair
superfluidity has previously been predicted in bilayer dipolar
systems where the particles are bound by an attractive interaction
between the layers \cite{Santos, Lew3, ripoll}. Though in bilayer systems, 
the state is a true molecular superfluid as $\Phi_i\ne 0$, whereas $\phi_i=0$ identically.
In the present system, in
spite of the particles interacting \emph{repulsively}, the pairs are
created due to the occupation-dependent tunneling terms in
Eq.~\eqref{Hamiltonian} (similar to \cite{Dutta}).  

To confirm that all these transitions are indeed quantum phase
transitions, we calculated -- for different chain lengths $N$ -- the
ground-state fidelity susceptibility \cite{Fidel3,Fidel1,Fidel2}
$\chi(d) = -\left.\frac{\partial^2 {\cal F}(d,\delta)}{\partial
\delta^2}\right|_{\delta=0}$, where ${\cal F}(d,\delta) = |\langle
\psi_0(d)|\psi_0(d+\delta)\rangle|$. Peaks in $\chi$ are efficient indicators of quantum
phase transitions. In  Fig.\ \ref{Fig2}(c), we present $\chi(d)$
for different chain sizes. There are three clear peaks at the
quantum phase transitions found from the correlation functions [as
presented in Fig.\ \ref{Fig2}(b)]. The positions of the transition
points (TPs) do not significantly depend on the number of sites,
especially for the 1st and 3rd TP. The middle peak in Fig.\ \ref{Fig2}(c) refers to the transitions
from checkerboard to antiferromagnetic superfluidity.  Moreover, the magnitude of the
fidelity susceptibility at all TPs increases with chain length,
which suggests that the transitions will survive in the
thermodynamic limit.

More insight into the properties of the observed phase comes from
the static structure factor, which is defined as $S(q) =
\frac{1}{N^2} \sum_{j,k=1}^N \mathrm{e}^{i q (j-k)} \left(\langle
\hat{n}_j \hat{n}_k\rangle - \langle \hat{n}_j\rangle\langle
\hat{n}_k\rangle\right)$, with $q=2\pi m/N$, $0\leq m\leq N-1$
integer.A peak in the structure factor at finite momentum points towards presence of
periodic density modulation in the systems. In Fig.\ \ref{Fig2}(d), we present $S(q)$ for a
half-filled system with $N=16$ sites. In the CB phase (between the
1st and 2nd TP), the dominant peak of $S(q)$ is at $q=\pi$, and its
magnitude is almost independent of system size. Above the
3rd TP, the system is in a phase where $\phi_i$ has an inverted
sign and $\Phi_i$ is large. This means that states where bosons
occur in pairs dominate (their contribution to the ground state is
about $95\%$). Since, due to the dipolar interactions, boson pairs
do not occupy neighboring sites, the system has some local
structure, leading to a predominant structure-factor peak at
$q=\frac{\pi}{2}$. The intermediate phase (between the 2nd and 3rd
TP) has interesting properties: the ground state of the finite
system deforms its structure stepwise, changing the dominant $q$
from $\pi$ to $\pi/2$ by one quantum $\Delta q = 2\pi/N$ at a time.
For $N=16$, this leads to three changes in the dominant $q$. Since
in an infinite system $q$ can take every value between $0$ and
$2\pi$, we expect in large chains a continuous change from the CB
with $q=\pi$ to the two-particle SF with $q=\pi/2$.
\begin{figure}
\includegraphics{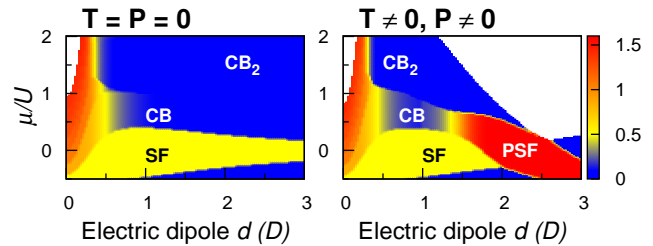}
\caption{ ED phase diagram without (left) and with (right)
taking into account $T$ and $P$. The color denotes the superfluidity fractions, $\phi_i$ and $\Phi_i$. Neglecting $T$ and $P$, for large
enough $d$ and $\mu$ the system is always in an insulating phase and
the average number of particles is a multiple of $1/2$. CB (CB$_2$)
denotes a checkerboard phase where sites with 0 and 1 (2) particles
alternate. Including the new terms, the insulating phases vanish for
large enough $d$, and a PSF appears. We truncate the Hilbert space
at a maximal occupation number of 4 particles per site. We exclude data points where the occupation number becomes too high (white region).}
\label{Fig4}
\end{figure}

\begin{figure}
\includegraphics{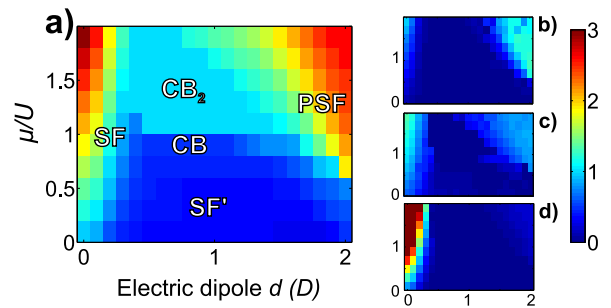}
\caption{Results for the BH Hamiltonian~\eqref{Hamiltonian} in a chain with $N=128$ using MERA with
$m=8$, revealing checkerboard (CB and CB$_2$) order, as well as superfluid (SF and SF') and pair-superfluid phases (PSF).
{\bf{(a)}} Mean occupation, {\bf{(b)}} mean SF order parameter, {\bf{(c)}} mean PSF order parameter, and {\bf{(d)}} mean NN density correlations.
\label{Fig5} }
\end{figure}

Finally, we analyze the influence of the additional terms $T$ and $P$ on the grand-canonical phase
diagram, where the particle number is not conserved.
For this, we add a chemical potential term $-\mu\sum_i \hat{n}_i$ to
Hamiltonian \eqref{Hamiltonian}.
In Fig.\ \ref{Fig4}, we present the
phase diagram as well as the average number of
particles per site for ED calculations of 4 sites with occupation truncated at 4 particles per site. When the additional terms $T$ and $P$ are large,
they destroy the CB phase, making place for a PSF.

To get a more detailed analysis of larger systems than tractable in ED, we have performed
a Multi-Scale-Entanglement-Renormalization-Ansatz (MERA) \cite{MERA,MERA_alg,mera} computation of the phase diagram
The MERA is a quasi-exact variational method that consists in postulating a tensor-network structure for the
low-energy states of Hamiltonian~\eqref{Hamiltonian}, 
which in particular yields especially good results in critical phases, where other methods such as DMRG are very costly \cite{MERA,MERA_alg}.

The results are presented in Figs.~\ref{Fig5}(a-d), where we show, averaged over the chain, the occupation $\langle n_i\rangle$, 
the SF order parameter $\langle a_i\rangle$, the PSF order parameter $\langle a_i a_i \rangle$, and NN density--density correlations $\langle n_i n_{i+1}\rangle$. 
The phase diagram extracted from these observables is sketched in Fig.~\ref{Fig5}(a).
At low $d$, there is a single-particle SF, which gives way to CB phases for $d\geq\mu$.
Increasing $d$, the system undergoes a
transition to a SF phase, where initially for a range of $\approx
0.2\mathrm{D}$ one-particle superfluidity dominates (similar to the ED results), and afterwards
pair superfluidity. At low $\mu$, we find a phase (SF') which has
additionally to SF order (i.e., a finite $\langle a_i\rangle$)
small nearest-neighbor density--density correlations.
Hence, it has a local structure where sites with
high and low occupation alternate. We checked that this phase is not due to phase separation.
The novel aspect of this is that in the usual extended BH model with soft-core interactions stable supersolidity appears only at the particle-doped region of the CB phase \cite{Sen, batr}.
For higher $\mu$ and $d \sim 1$, we get a CB of two
particles(CB$_2$ phase) in the filled site. This behavior is a
result of having low $U$ so that it is energetically favorable than
having one particle at each site. As already indicated by ED, the new terms $T$ and $P$ destroy CB order in favor of PSF phases, meaning that these terms cannot be neglected.
We also checked at few points in the phase space of the PSF region to look for the sign of the SF order parameter as a function of lattice sites and we found the alternating sign as seen in ED calculations.

\begin{figure} 
\includegraphics[scale=1.00, angle=0]{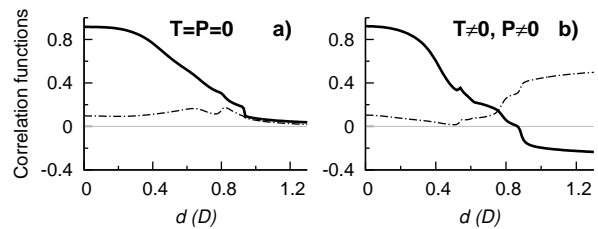}
\caption{The one-particle and two-particle correlation functions
$\phi_i$ (solid line) and $\Phi_i$ (dotted line) as a functions of dipole moment $d$ when the full dipolar interactions
are taken into account corresponds qualitatively to the calculations truncated at NNs (Fig.\ \ref{Fig2}). 
The large $\Phi_i$ and negative $\phi_i$ when terms $T$ and $P$ are taken into account indicate the break down of the CB phase to a PSF.
Calculations for ED at half-filling with $N=16$.}
\label{Fig6}
\end{figure}

To make better contact with experiment, we examine the disappearance of the CB pattern when the long-range part
of the full dipolar interactions is taken into account, i.e., we replace the NN term in Hamiltionian \eqref{Hamiltonian} 
with $\sum_{\{ij\},i\ne j} \frac{V}{|i-j|^3}\hat{n}_i\hat{n}_j$. Using ED at half-filling
for $N=16$, we find that qualitatively the phase diagram does not change much with
respect to our previous calculations with the simplified Hamiltonian \eqref{Hamiltonian} [compare Fig.\ \ref{Fig2}(b)]:
When the occupation-induced tunneling terms $T$ and $P$ are neglected, the CB phase remains stable for arbitrarily large $d$ [Fig.\ \ref{Fig6}(a)]. 
In contrast, when taking into account the tunneling terms $T$ and $P$ it disappears, making way for a PSF phase [Fig.\ \ref{Fig6}(b)].
This happens even at smaller $d$ than when truncating the interactions at NNs. Namely, the PSF phase appears for $d\ge 0.7\mathrm{D}$.
We also note that in Fig. \ref{Fig6}(b), there is a kink in $\phi_i$ around $d \sim 0.5\mathrm{D}$. This kink
corresponds to the appearance of a crystal like phase with modulation $|....200100200100....>$. A detailed discussion of this phase
is outside the scope of this paper. We have further checked that counter-intuitively PSF arises predominantly due to correlated tunneling $T$. 
Without this term PSF phase can not be reached for reasonable electric moments. 
We also note that for very low dipolar strength $\Phi_i$ has a small nonzero value. 
As seen in Figs.\ \ref{Fig2}(a), (b) and \ref{Fig6}(a), (b), a small but finite $\Phi_i$ is present as $d\rightarrow 0$ irrespective of
the presence of $T$ and $P$. This can be traced back to second-order processes due to $J$ which can also give rise to pair correlations
with small magnitude. 

In summary, we showed -- based on ED and MERA -- that commonly
neglected terms in the extended BH model for dipolar molecules in
optical lattices can lead to interesting new phenomena. We showed
for a particular choice of optical-lattice parameters that
occupation-dependent tunneling and pair tunneling (induced by
long-range dipolar interactions) destroy insulating checkerboard
phases for large enough electric moments $d$, leading to a novel
pair-SF phase. MERA results suggest also that a
supersolid phase could appear for $1/2$ filling even in the hole-doped case. 
Any presence of additional weak trapping potential
can result in shell like structures seen in usual BH model as long as local-density
approximation is valid. We note that, as our numerical calculations in carried out in one dimension,
the various superfluid correlations decay in a power law with distance. 
In this sense, the superfluid phases mention here will show quasi-long-range order in infinite systems.
Our calculations are done for parameters experimentally achievable in
the near future, and the changes to the phase diagram have to be
taken into account in the interpretation of future experiments with
dipolar molecules. 

This Letter was supported by the EU STREP NAME-QUAM, IP AQUTE, ERC Grant QUAGATUA, Spanish MICINN (FIS2008-00784 and Consolider QOIT), Caixa Manresa, and
Marie Curie project FP7-PEOPLE-2010-IIF ``ENGAGES'' 273524, and AAII-Hubbard.
T.S.\ acknowledges  hospitality  from  ICFO.

\end{document}


\title{Dipolar Molecules in Optical Lattices \\ Supplementary Material}
\author{Tomasz Sowi\'nski, Omjyoti Dutta, Philipp Hauke, Luca Tagliacozzo, Maciej Lewenstein}
\maketitle

\renewcommand{\theequation}{S\arabic{equation}}
\setcounter{equation}{0}
\renewcommand{\thefigure}{S\arabic{figure}}
\setcounter{figure}{0}

\subsection{Calculation of hopping terms T and P}
Here we describe the procedure to calculate the terms in the modified Hubbard model in Eq.~(1). 
First we find the lowest Bloch band for a single-particle moving in the potential $V_{\rm
latt}(\boldsymbol{r})=V_0 \left [ \sin^2 \frac{2\pi}{\lambda}x +
\sin^2 \frac{2\pi}{\lambda}y \right] + m\Omega_z^2 z^2/2$. From that,
we construct the  Wannier functions ${\cal W}^2_i(x,y)\mathrm{e}^{-\kappa z^2}$ localized at site $i$ \cite{Kohn}. 
By expanding the field operator in the Wannier basis, we derive the parameters for the Hubbard model. In
particular, the integrals used to calculate the
correlated hopping term $T$ and the pair-hopping term $P$ are:
\begin{eqnarray}
T&=&\int \int \mathrm{d}^3\boldsymbol{r}\mathrm{d}^3\boldsymbol{r'} {\cal W}^2_i(x,y)\mathrm{e}^{-\kappa z^2} {\cal V}(\boldsymbol{r}-\boldsymbol{r}') \nonumber\\
&\times& {\cal W}_i(x',y') {\cal W}_j(x',y')\mathrm{e}^{-\kappa z'^2} \nonumber\\
P&=&\frac{1}{2}\int \int \mathrm{d}^3\boldsymbol{r}\mathrm{d}^3\boldsymbol{r'} {\cal W}_i(x,y){\cal W}_j(x,y)\mathrm{e}^{-\kappa z^2} \nonumber\\
&\times& {\cal V}(\boldsymbol{r}-\boldsymbol{r}') {\cal W}_i(x',y') {\cal W}_j(x',y')\mathrm{e}^{-\kappa z'^2} \nonumber\\
\end{eqnarray}

\subsection{Description of MERA}
MERA is a variational method that consists in
postulating a tensor-network structure for the low-energy states of Hamiltonian. The tensor network $\mathcal{T}$ is i) built from elementary
tensors belonging to two different families, isometries ${\cal I}_i$
and disentanglers ${\cal D}_i$ that   are isometric,
\begin{equation}
{\cal I}_i {\cal I}_i^{\dagger}= \mathbb{I}; \quad {\cal D}_i {\cal D}_i^{\dagger}= \mathbb{I}\ ; \label{isometry}
\end{equation}
and ii) has a layered structure  ${\cal T}=\prod_i {\cal
T}_i$, such that each layer $\mathcal{T}_i$  performs
an ER transformation \cite{ER,ER_intro} from a lattice ${\cal L}_i$
with lattice spacing $b_i$ to a lattice ${\cal L}_{i+1}$ with
spacing $b_{i+1}=n b_i$.  Property ii) is at
the origin of the ability of the MERA ansatz to describe infinite
critical states with finite computational  resources. This is the
advantage of the MERA with respect to more traditional methods for
studying 1D chains such as, e.g., DMRG.
Symmetries of the Hamiltonian can be encoded in the structure of the
tensors. For example, in order to encode  translational invariant
states of chains with periodic boundary conditions, we use inside
each layer the same isometry and disentangler as many times as
required to complete the ER transformation from the lattice ${\cal
L}_i$ to the lattice ${\cal L}_{i+1}$. When all the isometries and disentaglers inside a given layer are chosen to be the same, the factor  $n$ not only characterize the blocking factor of the ER procedure (we talk about \emph{n to 1} MERA) but it also defines the size of the unit cell of the state. 

\begin{figure}
\centering
\includegraphics[width=0.22\textwidth]{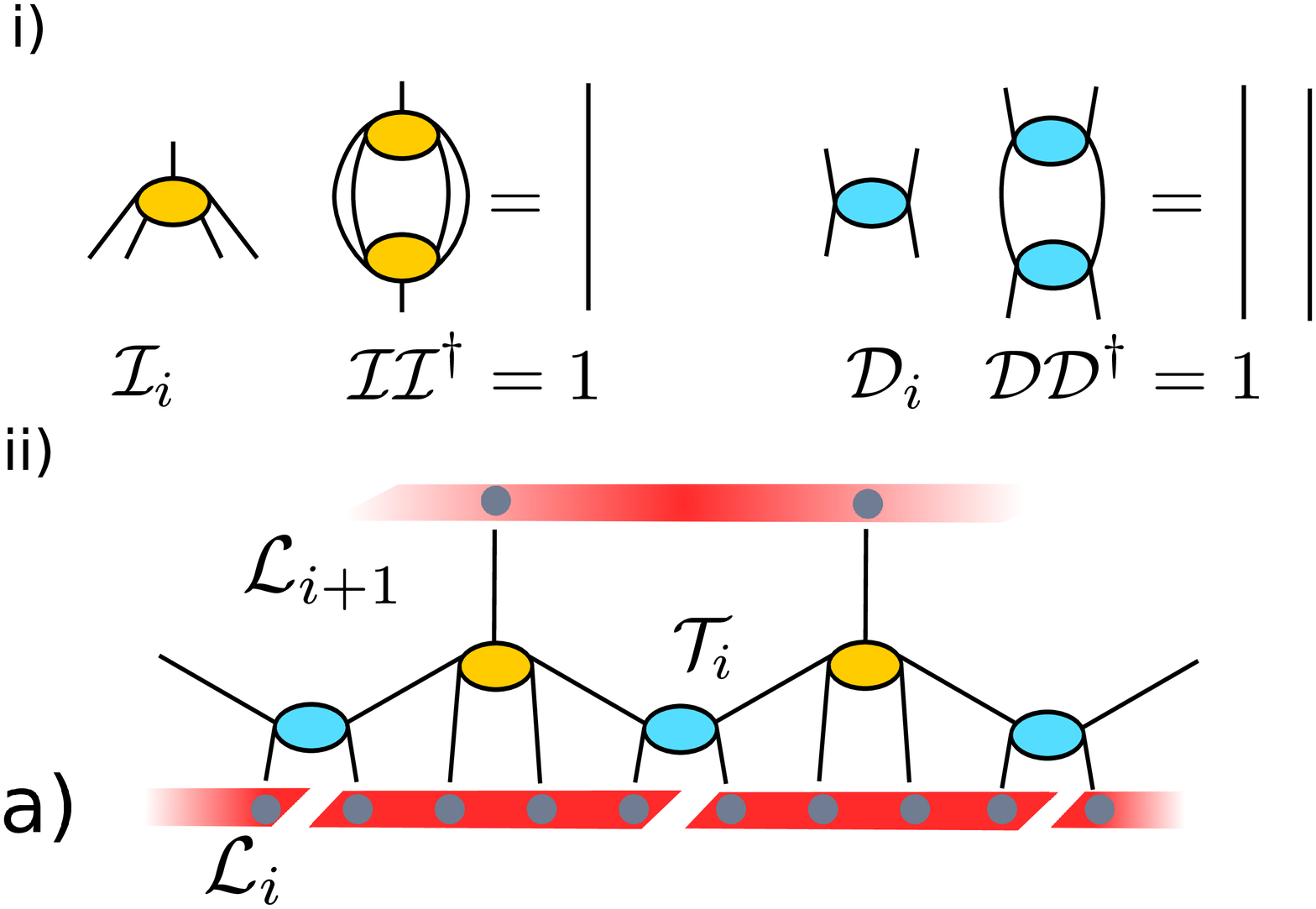}
\caption{ { Tensors ${\cal I}_i$ (isometries) and ${\cal
D}_i$ (disentanlgers) are represented by circles with trailing legs
representing their indices. Lines connecting two tensors represent
tensor contractions over the involved indices.  i) The tensors are
chosen such as to fullfill the isometry constraints defined in Eq.\
\eqref{isometry}. ii) A layer ${\cal T}_i$ of the  \emph{4 to 1} MERA tensor network ${\cal T}$ that maps operators and states defined on  a
lattice ${\cal L}_i$ with lattice spacing $b_i$ to operators and
states defined on a lattice ${\cal L}_{i+1}$ with lattice spacing $4
b_i$ \cite{MERA}.  
\label{Fig7} }}
\end{figure}

In the model we are considering  the presence of CB patterns in some parts of the phase diagram extracted
from ED suggests that we need  an ansatz that can naturally encode
at least a unit cell of two sites. This can be accomplished by a
\emph{2 to 1} MERA, i.e., by blocking two sites into one at each
step of the ER procedure. However, this MERA is
computationally more expensive than the   \emph{3 to 1} MERA. 
Unfortunately, the translationally invariant \emph{3 to 1} MERA does
not easily accomodate a CB pattern, whence we choose a
\emph{4 to 1} MERA that  both naturally accomodates the two-site
unit cell of a CB phase and reduces the computational cost
of the  \emph{2 to 1} MERA \cite{MERA_alg}. In Fig.\ \ref{Fig7}(a)
ii), we show a layer of the TN structure for the \emph{4 to 1} MERA
that we have used.
MERA has a refinement parameter $m$ larger values of which
provide more accurate results but imply larger simulation time,
since the complexity of the algorithm is ${\cal O}(m^5)$ in memory and ${\cal O}(m^8)$ in number of operations per iteration \cite{MERA_alg}, and modest values of
$m$ such as $m=8$ are often enough to get a correct qualitative picture of the model.